\documentclass[%
 reprint,
 amsmath,amssymb,
 aps,
floatfix,
]{revtex4-2}

\usepackage{graphicx}
\usepackage{dcolumn}
\usepackage{bm}
\usepackage{color}
\usepackage{appendix}
\usepackage[colorlinks=true,linkcolor=blue,anchorcolor=blue,citecolor=blue,urlcolor=blue,bookmarksnumbered]{hyperref}

\begin{document}


\title{Bright solitons in a spin-orbit-coupled dipolar Bose-Einstein condensate trapped within a double-lattice}
\thanks{Corresponding authors:\\
$\dag$ qinjieli@126.com \\
$\ddag$  lzhou@phy.ecnu.edu.cn\\
$\sharp$ phyzhxd@gmail.com}  %

\author{Qi Wang$^{1}$, Jieli Qin$^{2,\dag}$, Junjie Zhao$^{3}$,  Lu Qin$^{1}$, Yingying Zhang$^{1}$,  Lu Zhou$^{3,\ddag}$,  Xuejing Feng$^{1}$, Chunjie Yang$^{1}$, Zunlue Zhu$^{1}$, Wuming Liu$^{4}$ and  Xingdong Zhao$^{1,\sharp}$}

\affiliation{$^1$ School of Physics, Henan Normal University, Xinxiang, Henan 453007, China\\
$^2$ School of Physics and Electronic Engineering, Guangzhou University, Guangzhou, Guangdong  510006, China \\
$^3$ Department of Physics, School of Physics and Electronic Science, East China Normal University, Shanghai 200241, China \\
$^4$ Beijing National Laboratory for Condensed Matter Physics, Institute of Physics, Chinese Academy of Sciences, Beijing 100190, China
}

\date{\today}

\begin{abstract}
By effectively controlling the dipole-dipole interaction, we investigate the characteristics of the ground state of bright solitons in a spin-orbit coupled dipolar Bose-Einstein condensate. The dipolar atoms are trapped within a double-lattice which consists of a linear and a nonlinear  lattice. We derive the motion equations of the different spin components, taking the controlling mechanisms of the diolpe-dipole interaction into account. An analytical expression of dipole-dipole interaction is derived. By adjusting the dipole polarization angle, the dipole interaction can be adjusted from attraction to repulsion. On this basis, we study the generation and manipulation of the bright solitons using both the analytical variational method and numerical imaginary time evolution. The stability of the bright solitons is also analyzed and we map out the stability phase diagram. By adjusting the long-range dipole-dipole interaction, one can achieve manipulation of bright solitons in all aspects, including the existence, width, nodes, and stability. Considering the complexity of our system, our results will have enormous potential applications in quantum simulation of complex systems.

\end{abstract}

\maketitle


\section{\label{sec:1}Introduction}

Solitons are self-sustaining and localized wave packets that maintain their shape and velocity over long distances~\cite{Strecker2002,Nguyen2014,peng2019magnetized,Feng2023,Xue2023}. In recent years, solitons have been observed in various physical systems, including Bose-Einstein condensate (BEC) with spin-orbit coupling (SOC)~\cite{PhysRevLett.107.150403,PhysRevLett.109.095301,PhysRevA.87.023625,doi:10.1126/science.aaf6689}. The SOC-BEC systems are a type of ultracold atomic gas, where the atoms have an internal spin degree of freedom and are subjected to an external electromagnetic field~\cite{PhysRevA.80.033610,wei_spatiotemporal_2019}. These systems have been shown to support various types of solitons~\cite{PhysRevE.104.034214}, including bright solitons~\cite{PhysRevLett.90.040403,PhysRevA.82.013826,Young-S_2011}, dark solitons~\cite{PhysRevLett.83.5198,PhysRevA.77.033838,Frantzeskakis_2010} and so on~\cite{Qin:23,PhysRevA.102.063707}. These solitons have unique properties that make them suitable for applications such as quantum information processing~\cite{PhysRevA.95.053618,UTHAYAKUMAR2022128451} and quantum simulations~\cite{KUNDU2022128335}.

The dipolar Bose-Einstein condensates with large magnetic dipole moments observed in atomic sample such as $^{52}\mathrm{Cr}$~\cite{PhysRevLett.94.160401,PhysRevLett.95.150406,Lahaye2007}, $^{164}\mathrm{Dy}$~\cite{PhysRevLett.104.063001,PhysRevLett.107.190401}, and $^{168}\mathrm{Er}$~\cite{PhysRevLett.108.210401}, have opened new opportunities for the study of solitons dynamics in various contexts. Dipole-dipole interactions (DDI) are long-range and anisotropic, and both of the two features play an important role in the dynamic properties and spin dynamics of atoms~\cite{PhysRevA.78.043614,Lashkin_2009}. In condensed matter physics, DDI can arise from the intrinsic dipole moments of atoms or molecules, and can determine the physical properties of many materials. There has been growing interest in investigating the DDI on solitons in SOC-BEC~\cite{CHEN2020103304}. In this system, the DDI arises from the long-range magnetic and electric dipole moments of atoms, and can significantly affect the properties of the solitons~\cite{kengne2021spatiotemporal}.

When the BEC is trapped within optical lattice, the macroscopic quantum states of the ultra-cold atom have precise and controllable degrees of freedom, which makes an excellent platform for quantum simulation and one of the most important technologies that combines cold atomic physics with condensed matter physics. It is worth emphasizing specially that, Golam Ali Sekh et. al. discussed the effects of optical lattices on bright solitons in SOC-BEC, specifically on the geometric nodes and stability~\cite{sekh_effects_2021}. They focus on the effect of the nonlinear optical lattice that induced by a periodically modulated atomic scattering interaction but the DDI is ignored. In fact, it is well known that the long-range interactions will dominant the inter-site interactions as the atoms are separated in different lattices~\cite{PhysRevLett.113.013002}, though the collision interactions within the lattice still play an important role~\cite{RevModPhys.80.885,PhysRevLett.108.125301}. Our motivation is investigate the effect of DDI on the properties of bright solitons in the dipolar SOC-BEC trapped in double-lattice that consists of a linear optical lattice and a nonlinear optical lattice. By varying the dipole polarization axis, we found the DDI can play a major role and increased a controllable degree of freedom in this system.

This paper is organized as follows. In Sec.~\ref{sec:2}, after employing the standard method from three-dimension to quasi one-dimension, we introduce a Gross-Pitaevskii Equations (GPE) of the order parameter of two pseudo spin components. The analytic relationship between the DDI and the polarization angle and the linear dispersion relation are deduced.  In Sec.~\ref{sec:3}, following the Ritz optimization procedure based variational formulation of pairs of equations in the GPE, we study the effect of the system parameters on the bright soliton solutions, especially focus on the DDI. In Sec.~\ref{sec:4}, we employ the Vakhitov-Kolokolov stability criterion~
\cite{RevModPhys.83.247} to study the stability of the soliton solution obtained in Sec.~\ref{sec:3}. Section~\ref{sec:5} concludes and summarizes this work.

\section{\label{sec:2}Model}

\begin{figure}
\centering
\includegraphics[width=7.9cm,height=9.5cm]{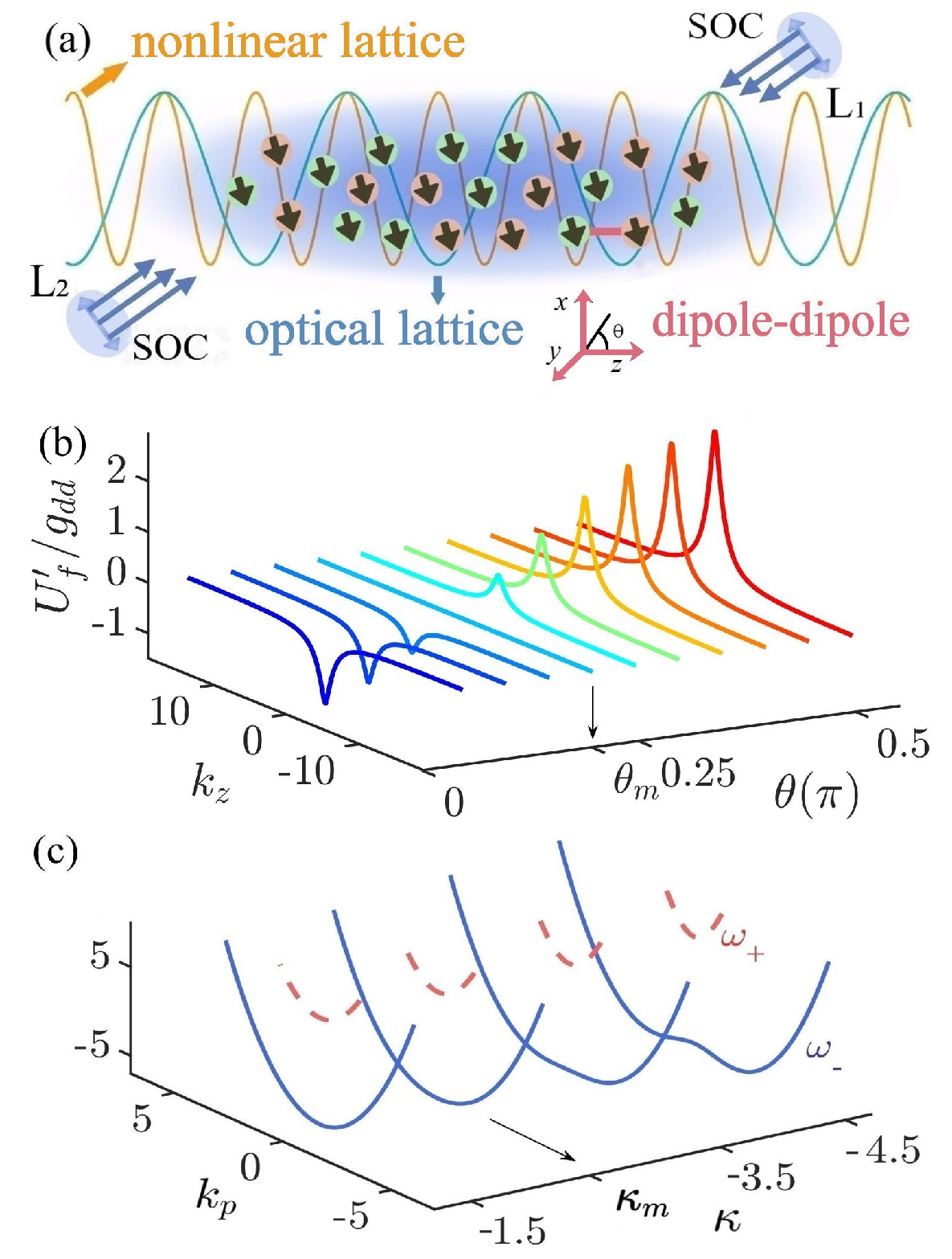}
\caption{(Color online)~(a)~The schematic diagram illustrates a dipolar BEC with SOC confined in a double-lattice, which consist of both a linear optical lattice (green line) and a nonlinear lattice (orange line). Two counter-propagating Raman lasers $L_1,L_2$ induce the SOC. The dipolar BEC is polarized in $x$-$z$ plane at an tunable angle of $\theta$ with respect to $x$-axis. (b) The effective DDI at different polarization angles.  When $\theta<\theta_m=\mathrm{arcsin}(\sqrt{3}/3)$, it is repulsive $(U_{f}^{'}/g_{dd}<0)$; while it is attractive $(U_{f}^{'}/g_{dd}>0)$ when $\theta>\theta_m$. (c) Linear dispersion relations of BEC at different SOC strengths $\kappa$. The upper branch (red dashed line) always has a single minimum. The lower branch (blue solid line) also has only a single minimum when $\kappa>\kappa_m$; However when $\kappa<\kappa_m$, it has two minimum.}
\label{fig1}
\end{figure}

We consider a quasi one-dimensional (1D) dipolar BEC system~\cite{Griesmaier_2007,PhysRevA.81.063616} with SOC realized by the Raman coupling scheme~ \cite{PhysRevA.105.053312,PhysRevE.103.052206,PhysRevA.81.025604}, which is shown in Fig.~\ref{fig1}(a). The system can be described by the following coupled nonlinear Schr\"{o}dinger equations~\cite{PhysRevA.92.033605}
\begin{small}
\begin{align}
i\partial_{t}\Phi_{j}  = &\left[-\partial_{z}^{2}/2+
(-1)^j \left(i\kappa\partial_{z}-\delta/2\right) + V\left(z\right)\right]\Phi_{j}+\Omega\Phi_{3-j}\nonumber \\
& -\left[ \gamma \left| \Phi_{j}\right|^{2} + \beta \left|\Phi_{3-j}\right|^{2} + D_{j} + D_{3-j} \right] \Phi_{j}, 
\label{eq:NSE1}
\end{align}
\end{small}
here, $j=1,\,2$ denotes the two spin components of the BEC,  $\kappa$ is the strength of SOC \cite{zhang2013tunable,Zhang_2017}, $\delta$ is the detuning \cite{PhysRevLett.109.115301}, $\Omega$ is the Rabi coupling frequency \cite{ravisankar2021effect}, $V\left(z\right)=V_{0}\cos\left(2\pi z/\lambda_{l}\right)$ is an optical lattice potential with strength $V_{0}$ and period $\lambda_{l}$, $\gamma$ and $\beta$ are two short-range contact interaction parameters, which can be tuned by the Feshbach resonance technique \cite{Inouye1998,PhysRevLett.96.170401,PhysRevLett.96.170401,HeZhangMing_2008,RevModPhys.82.1225,PhysRevA.97.060701}. In our work we assume the contact interactions are periodically modulated in space, thus the BEC also feels a nonlinear lattice, i.e., $\left\{ \gamma,\beta\right\} =\left\{ \gamma_{0} , \beta_{0}\right\} +\left\{ \gamma_{1} , \beta_{1}\right\} \cdot V_{n}\left(z\right)$ with $V_{n}\left(z\right) = \cos\left(2\pi z/ \lambda_{n} \right)$. 

For the last two terms on the right-hand side of Eq.~(\ref{eq:NSE1}), since we are considering a dipolar BEC \cite{adhikari2012dipolar}, we posit that the dipole moments, post-polarization, orient themselves parallel to the magnetic field direction and are restricted to the $x$-$z$ plane.  Consequently, the potential for DDI can be denoted as $U\left(\mathbf{r} ,\theta\right) = d^{2}/r^{5}\left[r^{2}-3(z \sin \theta+x \cos \theta)^{2}\right]$, where $d^{2}=\mu_{0} \mu^{2} / 4 \pi$, $\mu_{0}$ is the vacuum permeability \cite{PhysRevLett.85.1791}, $\mu$ is the dipole moment of the atom. Noted that the DDI is mathematically described by a convolution, it can be calculated efficiently by fast Fourier transform
\begin{align}
D_{j}=\mathcal F^{-1}\left\{ U_f\left(k_z,\theta\right) \cdot \mathcal F\left[\left|\Phi_{j}\left(z,t\right)\right|^{2}\right]\right\} ,
\label{D_{j}}
\end{align}
with $\mathcal F$ and $\mathcal F^{-1}$ standing for Fourier transform and its inverse respectively. The term $U_f\left(k_z,\theta\right)$ can be calculated analytically~\cite{PhysRevA.92.013633}.
\begin{align}
U_{f}\!\left(k_{z},\theta\right)\!=\!D_{d}\!\left\{ \!1\!-3\cos^{2}\!\theta\!\left[1\!-\!G\!\left(k_{z}\right)\right]/2\!-\!3\!\sin^{2}\!\theta G\!\left(k_{z}\right)\!\right\} .\!\label{eq:DDInteraction}
\end{align}
Here, $D_{d} = \mu_0 \mu^2/ 12 \pi L^2$ denotes the DDI strength and $L=\sqrt{\hbar/m \omega_\perp}$ represents the characteristic length of the system with the transverse trapping width $L$ of the optical lattice. It needs to be emphasized that the amplitude and the interacting range can be adjusted by tuning $\omega_\perp$, which was discussed in detail in our previous work~\cite{PhysRevB.76.214408}. Meanwhile, the sign of the DDI can be controlled by tuning the angle $(\pi/2-\theta)$ between dipole direction and $z$-axis~\cite{kawaguchi_spinor_2012,de_palo_formation_2022}, and $G\left(k_{z} \right) = k_{z} ^{2} \omega_{\bot} ^{2} \exp \left[k_{z}^{2}\omega_{\bot}^{2}\right]\Gamma\left(0,k_{z}^{2}\omega_{\bot}^{2}\right)$ with $\Gamma\left(\cdot\right)$ being the usual incomplete Gamma function. Easy to get $\lim _{k_{z} \rightarrow 0} U_{d d}\left(k_{z}, \theta\right)=U_{\infty}(\theta)=D_{d}\left(1-3 \sin ^{2} \theta\right)$. In Fig.~\ref{fig1}(b) we plot the relationship between $U_{f}^{'}=U_{f}-U_{\infty}$ and $\theta$, it is seen that the DDI is repulsive when $\left|\theta\right|<\theta_{m}=\mathrm{arcsin}\left(\sqrt{3}/3\right)$, while it is attractive when $\left|\theta\right|>\theta_{m}$, and at the critical angle $\theta=\theta_{m}$ the DDI disappears~\cite{PhysRevLett.115.023901}. By adjusting the transverse trapping width $\omega_\perp$ and the angle $\theta$, we obtain another parametric degree of the freedom to control the static and dynamical properties of the system, including the matter wave solitons.

In order to envisage a systematic study on the control mechanism of soliton in our double-lattice, we first focus on the relative importance of different parameters based on the linear dispersion relation~\cite{PhysRevLett.110.264101,Zhu2017}. By seeking plane wave solutions of Eq.~(\ref{eq:NSE1}), of the form $\Phi_{j}=\Phi_{0 j} \exp \left[i\left(k_{p} z-\omega t\right)\right]\left(\Phi_{0 j} \ll 1\right)$, we obtain the dispersion relation for the energy $\omega$ and momentum $k_p$,
\begin{align}
\omega_{ \pm}=\left(k_{p}^{2} / 2+V\right) \pm \sqrt{\kappa^{2} k_{p}^{2}+\Omega^{2}} .
\end{align}

It is obvious that the dispersion relation with SOC presents a double-branch structure as shown in Fig.~\ref{fig1}(c). The upper branch always has a single minimum while the lower branch contains two minimum as the strength of SOC exceeds a threshold $\kappa_m=-\sqrt{\Omega}$, though they are consistent below the threshold. For $\omega<\omega_{m}$, there exists a semi-infinite gap where linear modes do not propagate but matter-wave bright solitons with energies inside the gap can be found~\cite{PhysRevLett.92.230401,PhysRevLett.111.060402,PhysRevA.91.043629}. In the following, we apply analytical and numerical methods to analyze the characteristics of solitons in this region.  Especially, we will concern about the impact of the DDI.

\section{\label{sec:3} Method and result analysis}

\subsection{\label{sec:3a}Calculus of variations}

Firstly, it is an interesting curiosity to study how the chemical potential of each spin state changes in presence of the double lattice and interactions between atoms, because the chemical potentials lie below the $\omega_{m}$ in the semi-infinite gap determining the possibility of soliton existence. So we write the wave function in form $\Phi_{j} \left( z,t \right) = \phi_{j} \left( z \right) e^{-i \mu t}$, then the Lagrangian density is~\cite{PhysRevA.27.3135,ShenMing_2009}
\begin{align}
L=L_{1}+L_{2}+L_{12}+L_{nl} +L_{d},
\end{align}
with
\begin{subequations}
\begin{align}
&L_{j} = \left(\mu_{j}-V\right)\left|\phi_{j}\right|^{2} -\left|\phi_{j}^{'}\right|^{2}/2 - (-1)^j i \kappa \phi_{j} \phi_{j}^{*'},\\
&L_{12}= -\Omega\phi_{3-j}^{*}\phi_{j}  ,\\
&L_{nl} = 
\left[{\gamma_{1}} \left(\left| \phi_{1} \right|^{4} +\left| \phi_{2} \right|^{4} \right) /{2} +  \beta_{1} \left| \phi_{1} \right|^{2} \left| \phi_{2} \right|^{2}  \right]  V_{n} \left(z \right) \nonumber \\
&\quad \quad \quad +\beta_{0} \left|\phi_{1}\right|^{2}\left|\phi_{2}\right|^{2} + \gamma_{0} \left|\phi_{j}\right|^{4}/2 ,\\
&L_{d}= -\left( \left| \phi_{1} \right|^{2} + \left|\phi_{2}\right|^{2}\right)\left(D_{1}+D_{2}\right)/2 .&
\end{align}
\end{subequations}
here $\mu_{j}=\mu-\left(-1\right)^{j}\delta/2$ is the shifted chemical potential of the two spin components. The variational method offers a systematic and efficient approach for approximating solutions in intricate nonlinear systems, such as solitons. It facilitates the exploration of stability and existence aspects related to soliton solutions. By introducing variations within the trial function, one can assess the system's energy and stability characteristics. The trial wave function of the bright soliton will take the form~\cite{PhysRevA.87.013614}
\begin{align}
\phi_{j}\left(z\right)=A_{j}\exp\left[2i\left(-1\right)^{j}\pi z/J\right]\mathrm{sech}\left(z/a\right),
\label{brightsoliton}
\end{align}
where the variational parameters $A_{j}$ represent the amplitudes of the two spin components, and the variational parameter $a$ is the effective width of the soliton; while $J$ is a fixed parameter which can be taken from the optimization of the chemical potential.The trial wave function is normalized such that the atom numbers in the two spin components are $N_{i}=\int\left|\phi_{j}\right|^{2}dz=2aA_{i}^{2}$.
Inserting the trial wave function into Lagrangian density and integrating over the whole space, we obtain 
\begin{align}
\left\langle L\right\rangle =\left\langle L_{1}\right\rangle +\left\langle L_{2}\right\rangle +\left\langle L_{12}\right\rangle+\left\langle L_{nl}\right\rangle +\left\langle L_{d}\right\rangle ,
\end{align}
with
\begin{subequations}
\begin{align}
&\left\langle L_{j}\right\rangle =  N_{j}\left[\mu_{j}+W - a\pi^{2}  V_{0} \operatorname{csch} \left(a \pi^{2} / \lambda_{l} \right)/\lambda_{l}\right],\\
&\left\langle L_{12}\right\rangle=  -2 a \pi^{2} \Omega \sqrt{N_{1} \! N_{2}} \operatorname{csch}\left(2 a \pi^{2}\! /\!J\right)/J, \\
&\left\langle L_{n l}\right\rangle = N_1 N_2 \left(2R \beta_{1} +\beta_0 /3a  \right)  \nonumber \\
& \quad\quad\quad \,\  + R\gamma_1 \left(N_{1}^2+ N_{2}^2 \right) + (N_1 + N_2) \gamma_0/6a ,\\
&\left\langle L_{d}\right\rangle =  -{\pi a^{2}} \left( N_{1} + N_{2} \right)^{2} \Theta \left(a,\theta\right) /16,
\end{align}
\end{subequations} 
here $W=-a^{2}/6+2\pi /{J}(\kappa-\pi/J)$ and $R={\left(a^{2} \pi^{4}  +\pi^{2}\lambda_{n}^{2} \right)}$$\operatorname{csch} \left({a \pi^{2}}/ {\lambda_{n}}\right) /{6 \lambda_{n}^{3}}$. In this step all the integrals can be calculated analytically, except that in $\left\langle L_{d}\right\rangle $ the function $\Theta\left(a,\theta\right)=\int\mathrm{csch}^{2}\left({\pi ak_{z}}/ {2}\right) U_{f} \left(k_{z},  \theta \right)k_{z}^{2}dk_{z}$ needs a numerical treatment due to the complexity of DDI. Employing the Ritz optimization procedure, from the condition $\partial\left\langle L\right\rangle /\partial N_{j}=0$, we can determine the chemical potentials 
\begin{figure}[b]  
\centering
\includegraphics[width=8.5cm,height=3.2cm]{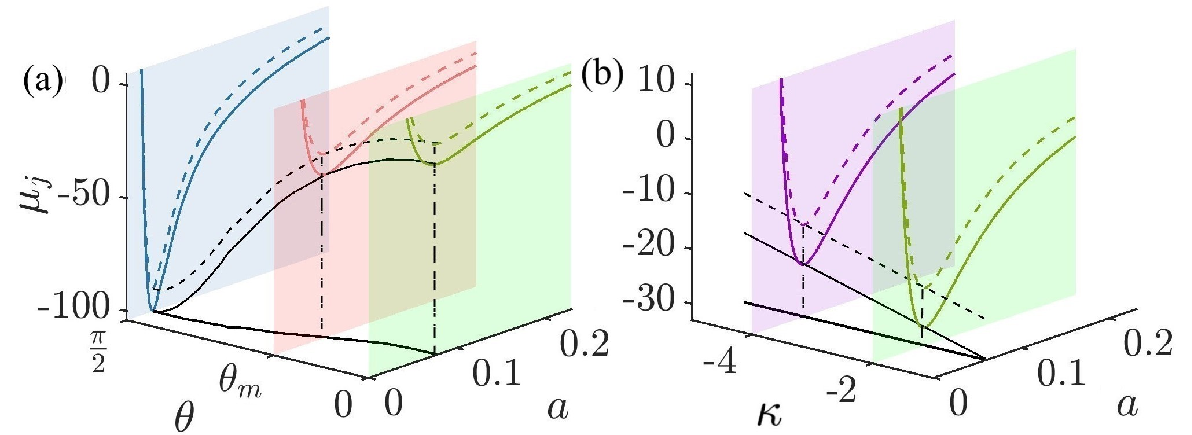}
\caption{(Color online)~The relationships between chemical potential $\mu_j$ and soliton width $a$ under different (a) DDI strengths and (b) different SOC strengths. The colored solid line represents $\mu_1$ while the colored dashed line represents the $\mu_{2}$. In (a), the red, green and blue lines correspond to polarization angles $\theta=\mathrm{arcsin}(\sqrt{3}/3)$,~$0$,~$\pi/2$ respectively. The SOC strength is $\kappa = -2.0$. In (b), the green and purple lines correspond to SOC strengths $\kappa = -3.0$ and $-1.0$. The polarization angle is $\theta =0$. In both (a) and (b), the black solid and dashed lines represent the minimum of the chemical potential. Other parameters are $D_d=3.0$, $\gamma_0 = 2$, $ \beta_0 = 2$, $V_0 =-4$, $ \gamma_1 =0.5$, $ \beta_1 = 0.5$, $N_1 = 5.0$, $ N_2 = 1.5$, $\lambda_l = 0.65$, $ \lambda_n = \lambda_l/2$, $J = 2$ and $\Omega = 6$.} 
\label{fig2}
\end{figure}
\begin{align}
\mu_j \!&=\!  a\pi^{2}  V_{0} \operatorname{csch} \left(a \pi^{2} / \lambda_{l}  \right)/\lambda_{l} \!-\! 2R\left(\beta_{1} N_{3-j} \!+\! \gamma_1 N_{j} \right) \!-\! W \nonumber \\
&\quad+\!{\pi a^{2}}\left(N_{1} \!+\! N_{3-j}\right) \Theta\left(a,\theta\right) /{8} \!-\! \beta_{0} N_{3-j}/{3 a} \!-\! \gamma_{0}/{6 a}  \nonumber \\
&\quad+\!a \pi^{2} \Omega \sqrt{N_{3-j}} \operatorname{csch}\left(2 a \pi^{2} /J\right)/\sqrt{N_{j}} J .
\label{mumujj}
\end{align}

Except the structure parameters of the soliton, the optical lattice and DDI also have contributions to the chemical potential, the former was details studied in depth in the article by Golam Ali Sekh~\cite{sekh_effects_2021} and the latter has not been investigated in this system so far. Usually, a negative minimum of the chemical potential indicates the existence of a self-bounded state, i.e., a bright soliton. 

\subsection{\label{sec:3b} Influence on chemical potential}

Based on this idea, we examine the relationship between chemical potential $\mu_{j}$ and effective soliton width $a$. We are mostly concerned with the effect of DDI on the solitons, therefore we plot the $\mu_{j}$-$a$ curves for different DDI in Fig.~\ref{fig2}(a). For the red curves, we set $\theta=\theta_{m}$, i.e., the DDI vanishes, and we do observe a negative minimum point on the curve, this means a bright soliton can exist without the DDI (in fact solitons induced by the short range contact interaction have been widely studied). For the blue curves, the DDI is set to be attractive with polarization angle $\theta=9\pi/20$, comparing it with the red curves, we see that the minimum chemical potential is reduced, at the same time the effective soliton width $a$ corresponding to the minimum chemical potential also becomes smaller. The situation is slightly different for the SOC, the minimum chemical potentials increases when the strength of the SOC increases. However, the effective soliton width is independent with this change as shown in Fig.~\ref{fig2}(b).
\begin{figure}[t]
\includegraphics[width=8.4cm,height=6.4cm]{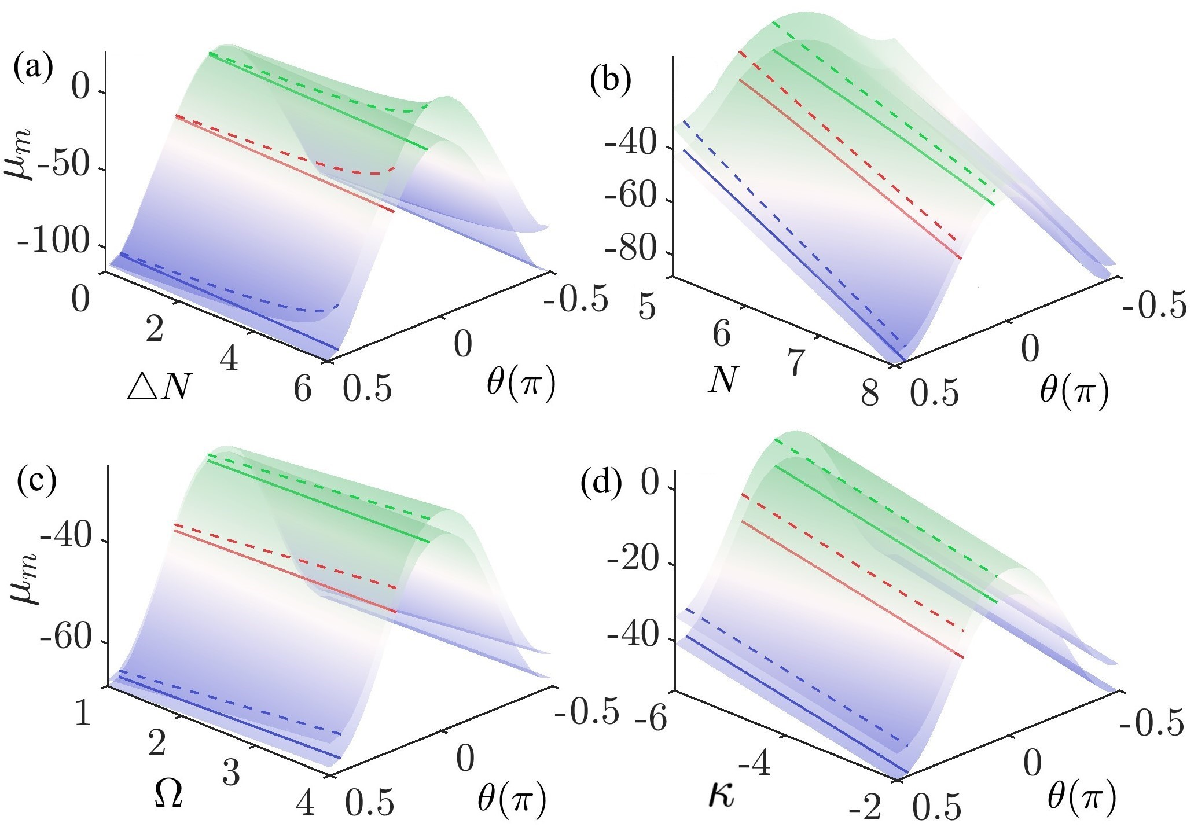}
\caption{(Color online) The relationships between minimum chemical potential $\mu_j$ and atomic difference $\Delta N$ (a), total atomic number $N$ (b), Rabi coupling frequency $\Omega$ (c), SOC strength $\kappa$ (d) at different polarization angles $\theta$. The upper surface represents the minimum chemical potential ($\mu_{2m}$) corresponding to $\phi_2$, and the lower surface represents the minimum chemical potential ($\mu_{1m}$) corresponding to $\phi_1$. The lines on the surfaces correspond to polarization angles $\theta=\pi/20$ (green), $\mathrm{arcsin}(\sqrt{3}/{3})$ (red), and $9\pi/20$ (blue). The solid line is $\mu_1$, the dashed line is $\mu_2$. Other parameters are the same as the green line in Fig.~\ref{fig2}, if they are not explicitly shown on the graphs.}
\label{fig3}
\end{figure}
From Eq.~(\ref{mumujj}), the atomic numbers in the two spin components have contribution in determining the chemical potential, though they are fixed in the above discussion. In fact, the imbalance in the number of particles ($N_1 \ne N_2$) can be induced by the spontaneous oscillations between the two pseudo-spin components, which determines the statistical method used to describe the system. From another perspective, the atomic difference between two spin states $\Delta N$ can be used to define the magnetization of the system
\begin{align}
M= \Delta N/N,
\end{align}
with
\begin{align}
N_1 = (N\pm \Delta N)/2  \quad \text{and} \quad N_2 = (N\mp \Delta N)/2.
\end{align}

We plot $\mu_m$ versvs $\Delta N$ for a fixed value of $N$ in Fig.~\ref{fig3}(a) and $\mu_m$ versus $N$ for a fixed $\Delta N$ in Fig.~\ref{fig3}(b). It is obvious that the curved surface for $\mu_m$ bifurcates when the magnetization of the system occurs, and the separating interval between the curved surfaces increases with $\Delta N$. On the contrary, for a value of $\Delta N$, the chemical potential of the two spin states decreases gradually as $N$ increases, which corresponds to the decreased magnetization, and the separating interval also decreases as shown in~\ref{fig3}(b). The above analysis implies that the magnetization of the system and the difference of the chemical potential of the two spin states are positively correlated. It is worth emphasizing that the DDI has effect on the difference in chemical potential, but its contribution to their magnitude is significant.
\begin{figure}[b]
\centering
\includegraphics[width=8.3cm,height=3.2cm]{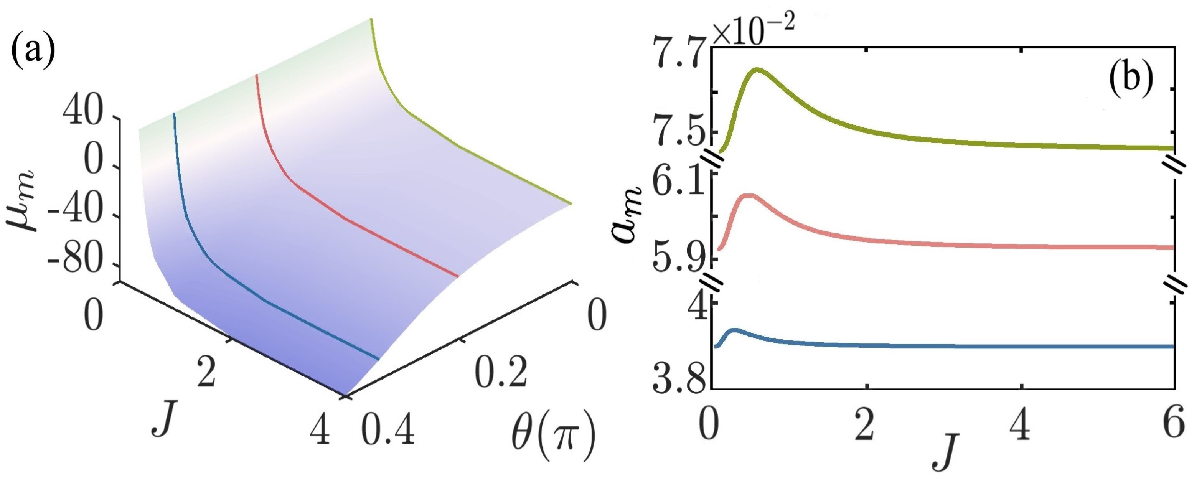}
\caption{(Color online) (a) The relationship between minimum chemical potential $\mu_1$ and variational parameter $J$ at different polarization angles. (b) The corresponding effective soliton width $a_m$ as a function of $J$. The green, red and blue lines correspond to $\theta=0, \mathrm{arcsin}(\sqrt{3}/3), \pi/3$ respectively. Other parameters are the same as the green solid line in Fig.~\ref{fig2}. }
\label{fig4}
\end{figure}

\begin{figure*}
\centering
\includegraphics[width=16cm,height=7.8cm]{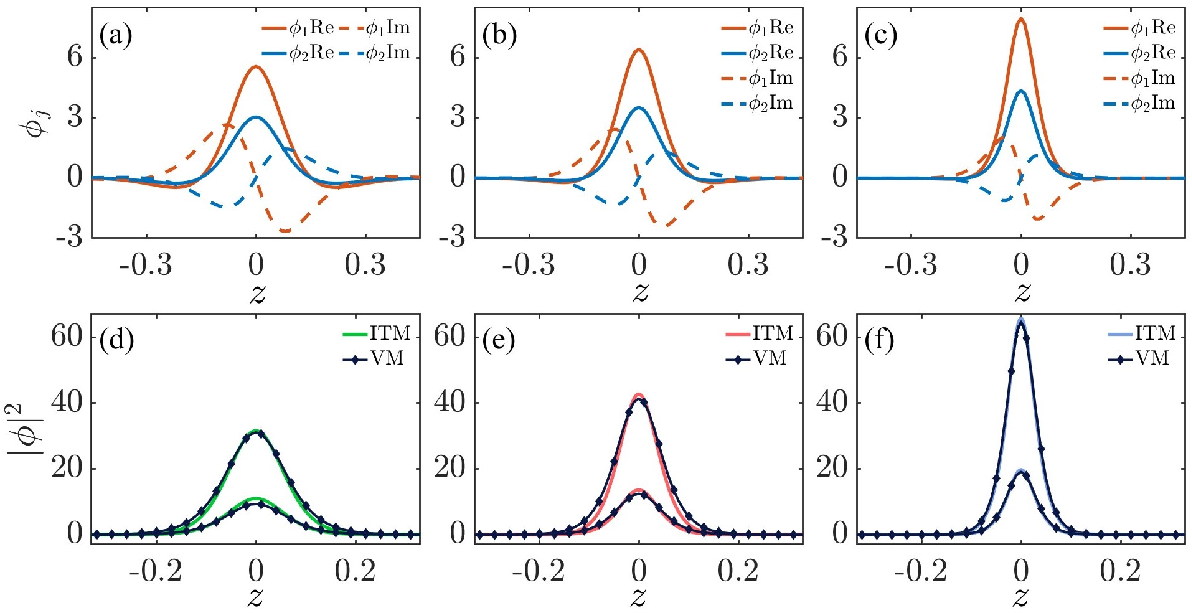}
\caption{(Color online) The wave functions (a)-(c) and density profiles (d)-(f) of some typical solitons at different polarization angles $\theta=0$ (a)(d), $\mathrm{arcsin}(\sqrt{3}/3)$ (b)(e), $\pi/3$ (c)(f). In (a)-(c), the solid and dashed line styles represent the real and imaginary parts of the wave function, and the red and blue line colors represent the two spin components. In (d)-(f), the dots represent the variational method (VM) results, while the solid line is obtained using the imaginary time method (ITM). Other parameters are the same as those in Fig.~\ref{fig2}.}
\label{fig5}
\end{figure*}

In addition to atomic numbers, there are many adjustable parameters for us to choose from in this system. To check the effect of the different parameters, we display the chemical potential $\mu_m$ with the variation of the Rabi frequency $\Omega$ and the SOC strength $\kappa$ in Fig.~\ref{fig3}(c) and (d) respectively. It is shown that the effects of $\Omega$ and  $\kappa$ on the chemical potential $\mu_m$ are obviously different. Increasing the Rabi frequency increases the fluctuation of atomic numbers and ultimately induces an increase in the minimum value of chemical potential. Moreover, the effect of $\Omega$ depends on the initial population distribution, which leads to the increasing separation of the two chemical potential surfaces as $\Omega$ increases. Conversely, increasing the SOC strength $\kappa$ decreases the minimum value of chemical potential, and the two spin states share the similar trend which results in unchanged separation of the two chemical potential surfaces.  

Except the adjustable parameters of the system discussed above, Eq.~(\ref{mumujj}) consists of some other parameters that represent the characteristics of the matter wave-bright solitons. For instance, the parameters $J$ and $a$ come from the proposed trial solution of Eq.~(\ref{brightsoliton}). In Fig.~\ref{fig4}, we plot soliton width $a_m$ versus $J$ for different polarized angle $\theta$, As $J$ increases, the chemical potential $\mu_m$ gradually decreases and undergoes a transition from positive to negative(a) , and $a_m$ will first increases and then gradually decreases(b). There always exists a maximum point no matter how the dipole interaction is adjusted. This tells us that it is possible to prepare the weakly or strongly bound state by adjusting the parameter $J$ and tuning the DDI which is more feasible. It is worth mentioning that this process can be achieved by simultaneously adjusting the nonlinear optical lattice as proposed in Ref.~\cite{sekh_effects_2021}. Our method is fundamentally different from theirs. Here the long-range interaction is tuned and any assumptions about lattice potential are not broken.

Based on the above discussed parameters, we can obtain clear images of bright soliton shapes, which are displayed in Fig.~\ref{fig5}. Fig.~\ref{fig5}(a)-(c) shows the spatial variation of real and imaginary parts of $\phi_j(j=1,2)$ with different polarized angle $\theta$. It is obvious that the amplitudes of both the real and imaginary parts of $\phi_1$ is large than those of $\phi_2$. Regardless of the adjustment of the DDI, the imaginary parts of $\phi_j$ always exhibit nodes, but the nodes of the real parts would disappear as shown in  Fig.~\ref{fig5}(c). When we adjust the dipole interaction from repulsion to zero and then to attraction, the widths of the entire wave packet will gradually narrow. This means that by adjusting the dipole interaction, we can effectively adjust the shape of bright solitons. To verify the effectiveness of the variational method, we also employ the imaginary time evolution method to find soliton solutions under the same parameters, the comparison of the two methods are shown in Fig.~\ref{fig5}(d)-(f). The results obtained by the two methods have good consistency. 

In sum, we obtained the characteristics of bright solitons through the variational method, and we have achieved effective control of the bright solitons by adjusting system parameters, especially the DDI. If we want to apply the results here to the quantum simulation process, we must pay attention to the stability or lifetime of the solitons generated here for effective observation in experiments. 

\begin{figure*}[htb]
\centering
\includegraphics[width=16cm,height=7.8cm]{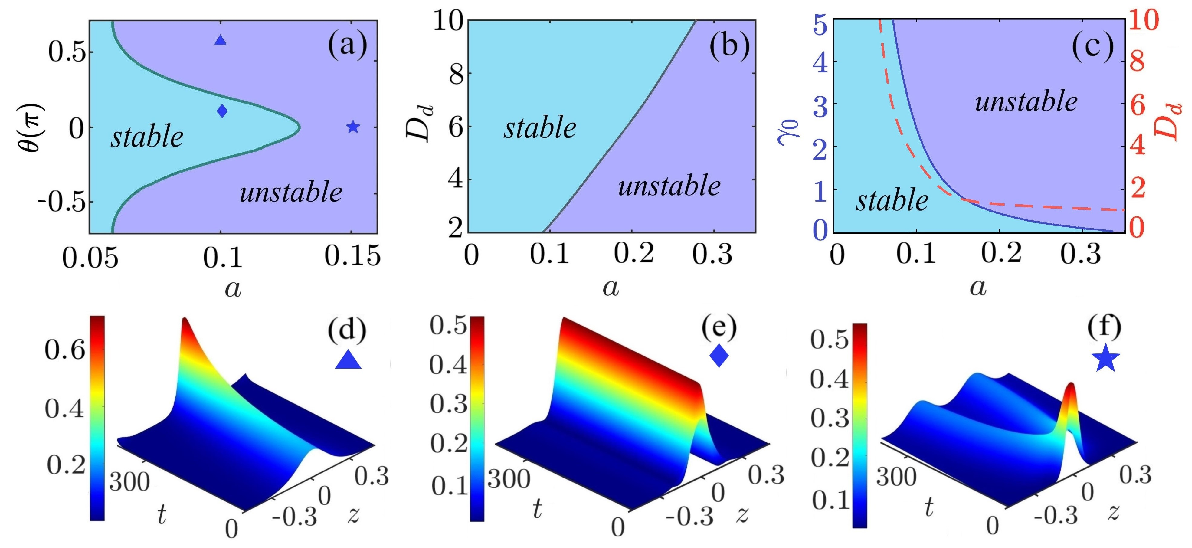}
\caption{(Color online) (a)-(c) Stability phase diagrams. The blue part is the stable region, and the purple part is the unstable region. (a) Indicates that the soliton stable region varies with the $\theta$ a between the DDI and the $x$-axis. (b) Represents the variation of the soliton stability region with the variation of the $D_d$ coefficient at $\theta = 0$. (c) Shows the variation of soliton stability region tuned by Feshbach resonance technique under the condition of solid line ($\theta =\arcsin \left(\sqrt{3}/3\right)$, ~$\gamma_0 =\beta_0$). The dashed line ($\theta =\pi/3$, ~$\gamma_0=\beta_0=0$) shows that the soliton's stable region changes with the variation of the  coefficient $D_d$ under attractive DDI. (d)-(f) Select the values of the unstable region and the stable region in (a) respectively for time evolution. (d) $a=0.1$,~$\theta = \pi/3 $ in the unstable region is selected and corresponds to the triangle in (a) above; (e) $a=0.1$, ~$\theta =\pi / 10$ in the stable region is selected and corresponds to the lozenge in (a) above. (f) $a=0.15$, ~$\theta = 0 $ in the unstable region is selected and corresponds to the pentagram in (a) above. If not specifically mentioned in the subfigure, the parameters are as follows $\theta =0$,~$Dd=3.0$,~$J = 2$,~$ \beta_0 = \gamma_0 = 2$,~$\lambda_l = 0.65$,~$\lambda_n = \lambda_l /2$,~$V_0 =-4$,~$\beta_1 = \gamma_1 = 0.5$,~$\Omega = 6$,~$ N_1 = 5$,~$ N_2=1.5$,~$\kappa=-2.0.$}
\label{fig6}
\end{figure*}

\subsection{\label{sec:4} Stability region}

Besides changing the soliton profile, the DDI would also affect the stability of the solitons. This can be studied by the well-known Vakhitov-Kolokolov (VK) criterion~\cite{PhysRevB.76.214408,RevModPhys.83.247,PhysRevA.99.043832}, which relies on the sign of quantity $\partial N_{j}/\partial\mu_{j}$ --- if $\partial N_{j}/\partial\mu_{j}>0$, the corresponding soliton is stable; otherwise it is unstable. To obtain this quantity, we firstly determine $N_{i}$ by optimization condition $\partial\left\langle L\right\rangle /\partial a=0$,
which gives
\begin{equation}
N_{1}=\left( T_{1}+T_{2}\right)/\left(T_{3}+T_{4}+T_{5}\right),
\label{NNjj}
\end{equation}
with 
\begin{subequations}
\begin{align}
T_1=& -{24 \pi^2 \Omega  \sqrt{s}} \left[J-2 a \pi^{2} \text{coth}\left({2 a \pi^{2}}/ {J}\right) \right] \nonumber /{J^{2}} \\
& \times  \text{csch}\left({2 a \pi^{2}}/{J}\right) ,\\
T_2=&\left\{{6 \pi^{2} V_0 } \left[  a \pi^{2} \text{coth} \left({a \pi^{2}} / {\lambda_l} \right)- \lambda_l\right]\text{csch} \left({a \pi^{2}} / {\lambda_l} \right)\right /{\lambda_l^{2}} \nonumber\\
 &\left.+ {2} /{a^{3}} \right\}(1+s),\\
T_3 =&\left(2 \beta_0 s+\gamma_0+s^{2} \gamma_0 \right) /{a^{2}} ,\\
T_4=&{\pi^{4}} \left[-2 a \lambda_n + \left(\lambda_n^{2}+a^{2} \pi^{2}\right) \text{coth}\left({a \pi^{2}} / {\lambda_n} \right) \right] /{\lambda_n^{4}} \nonumber \\
& \times  \left(2 \beta_1 s+\gamma_1+s^{2} \gamma_1\right)  \text{csch}\left({a \pi^{2}} / {\lambda_n}\right) , \\
T_5 =& {3 a \pi^2 (1+s)^{2}}\int  k_z^{2}  \left[2 - a \pi k_z \operatorname{coth}\left({a \pi k_z }/ {2}\right)\right]/8 \nonumber \\
& \times \operatorname{csch} \left({a \pi  k_z} / {2} \right)^2 U_f(k_z , \theta ) dk_z ,
\end{align}
\end{subequations}
here $s=N_{2}/N_{1}$ is a parameter describing the ratio between atom numbers in the two spin components. Now $N_{i}$ and $\mu_{j}$ can both be expressed as a function of soliton width $a$ {[}Eq.~(\ref{NNjj}) and Eq.~(\ref{mumujj}){]}, the VK criterion can be calculated by
\begin{align}
\frac{\partial N_{j}}{\partial\mu_{j}}=\frac{\partial N_{j}}{\partial a}\frac{\partial a}{\partial\mu_{j}}.
\end{align}

The stability phase diagrams determined by the VK criterion are plotted in Fig.~\ref{fig6}. In Fig.~\ref{fig6}(a), the DDI is tuned from attractive ($\theta=\pm\pi/2$) to repulsive ($\theta=0$), and it is found that under the attractive DDI only the solitons with very small effective width $a$ are stable; While under repulsive DDI the stable region extends to a larger value of $a$. In Fig.~\ref{fig6}(b), the attractive/repulsive property of the DDI is fixed, only its strength $D_{d}$ varies, we found that as $D_{d}$ increases, the repulsive/attractive interaction can gradually enlarge/reduce the stable region. In Fig.~\ref{fig6}(c), the DDI is fixed. By adjusting the modulation intensity of Feshbash, we found that the stable region can be gradually reduced as $\gamma_0$ increases. 

In order to make a comparation between the long-range DDI and the short-range collision interaction, we tune the angle $\theta$ to $\pi/3$ as shown in Fig.~\ref{fig6} (c). It is obvious that the soliton with small width becomes unstable when the amplitude of the DDI increases. However, when the width $a$ increases the soliton dominated by the long-range DDI exhibits stronger stability than the soliton dominated by the short-range interaction. In other words, DDI has a more significant impact on solitons with larger widths.      

We also confirmed the stability property of the soliton by directly simulating the dynamical Eq.~(\ref{eq:NSE1}), the results are shown in Fig.~\ref{fig6}~\cite{PhysRevA.106.043307,PhysRevLett.115.180402,gangsoliton2006,PhysRevLett.104.073603}. In Fig.~\ref{fig6} (d), the initial soliton state lies in the unstable region [represented by $\blacktriangle$ in Fig.~\ref{fig6} (a)], and we see that during the evolution the soliton wavepacket does not keep its initial shape, indeed the attractive DDI leads to a shrinking of the wavepacket. In Fig.~\ref{fig6} (e), the initial soliton state lies in the stable region [represented by $\blacklozenge$ in Fig.~\ref{fig6}(a)], this time we do observe a stable evolution of the soliton. At last, we also examined the evolution of unstable soliton under repulsive DDI [represented by $\bigstar$ in Fig.~\ref{fig6}(a)], the numerical result show that such a soliton firstly endures a spatial spreading, and then it spontaneously splits into two individual soliton wavepackets with halved atom number.


\section{\label{sec:5}CONCLUSION}

In summary, we studied bright solitons in the system of spin-orbit coupled dipolar BEC subjected to both a linear and a nonlinear lattice.  In the system, DDI can be tuned to be either attractive or repulsive by adjusting the polarization angle. We found that in such a way both the profile and stability region of the solitons can be engineered --- the repulsive DDI will broaden the soliton and enlarge the stability region, while the attractive DDI tends to narrow the soliton and reduce the stability region. We also studied the instability dynamics by numerically simulating the nonlinear Schr\"{o}dinger equations. It is found that attractive DDI leads to a shrinking of the soliton, while the repulsive DDI results in spreading and splitting of the soliton.

Overall, this research furnishes crucial insights into matter wave dynamics within Bose-Einstein condensates, underscoring the pivotal regulatory function of DDIs therein. Furthermore, it unveils alternative methodologies for manipulating solitons within spin-orbit coupled Bose-Einstein condensates, presenting a substantive basis for subsequent investigations and applications.

\hspace*{\fill}

\textbf{Acknowledgments:}
This work was supported by the National Key Research and Development Program of China (2021YFA1400900, 2021YFA0718300, 2021YFA1400243),  National Natural Science Foundation of China (12074105, 12104135, 61835013, 12074120, 11904063).

\section*{APPENDIX: Three-dimensional to Quasi-one-dimensional}
\setcounter{equation}{0}
\renewcommand\theequation{S\arabic{equation}}
The Hamiltonian of a dipolar SOC-BEC within double-lattice under the mean field approximation

\begin{equation}
H =H_0+H_{s} + V(\vec{r}) + H_{nl} + H_{d}.
\end{equation}
Here, $H_0$ is the Hamiltonian of a single particle, $H_s$ is the Hamiltonian of SOC, $V(\vec{r})$ is the external potential field, and $H_{nl}$ comes from the nonlinear optical lattice. $H_d$ is the Hamiltonian of dipole dipole interaction.
\begin{subequations}
\begin{align}
H_{0}&=\frac{{\vec{p}}^{2}}{2 m},\\
H_{s}&=\kappa^{\prime}{p_x}  {\sigma_z} + \Omega^{\prime}\sigma_{x}\hbar,
\end{align}
\end{subequations}
where $\Omega^{\prime}$ stands for Rabi frequency and $\sigma_{x,z}$ the Pauli spin matrices. The Hamiltonian implication of short-range interactions is provided by

\begin{equation}
H_{nl}=
\begin{pmatrix}  \gamma|\Psi_1|^2+\beta|\Psi_2|^2 & 0 \\
0 &\gamma|\Psi_2|^2+\beta|\Psi_1|^2 \end{pmatrix},
\end{equation}
here $\gamma$ and $\beta$ stand for intra- and inter-atomic interactions. The contribution of the DDI is given by
\begin{align}
H_{d}=\sum_{i,j}^{1,2} \int \Psi_{i}^{*}\left(\vec{r}, t\right) U \left(\vec{r}-\vec{r}^{\prime}\right) \Psi_{j}\left(\vec{r}^{\prime}, t\right) d^3 \vec{r}^{\prime},
\end{align}
where the dipole interaction potential can be expressed as
\begin{align}	
U(\vec{r} - \vec{r}^{\prime}) = \frac{\mu_0 d^2}{4\pi}\frac{1-3 \cos ^{2} \alpha}{\left|\vec{r} - \vec{r}^{\prime} \right|^{3}},
\end{align}
where $\mu_0$ is the permeability of vacuum, $d$ is magnitude of the dipole vector, and $\alpha$ the crossing angle between $\vec{d}$ and $\vec{r} - \vec{r^{\prime}}$. We assume that the dipole moment is polarized in the direction of the magnetic field and is restricted to the $x$-$z$ plane, so $\vec{d}=d[\sin (\theta) \hat{\mathbf{z}}+\cos (\theta) \hat{\mathbf{x}}]$. When capturing potential field $V(\vec{r})=m(\omega_x^2 x^2 + \omega_y^2 y^2) /2+ V_0 \cos(2\lambda_l z)$, and we take $\omega_\perp = \omega_x = \omega_y$, the wave function of the system is changed from three dimensions to quasi-one dimension by a standardized simplification method. Let's rewrite the wave function of the system
\begin{equation}
\begin{aligned}	
\Psi_j(\vec{r},t)=&\Phi_x(x)\Phi_y(y)\Phi_{j}(z,t),\\
\Phi_x=&\frac{1}{\sqrt[4]{\pi L^2}}e^{-\frac{x^2}{2L^2}},\\ \Phi_y=&\frac{1}{\sqrt[4]{\pi L^2}}e^{-\frac{y^2}{2L^2}},
\end{aligned}
\end{equation}
where $L=\sqrt{ {\hbar}/{m\omega_\perp}}$ represents the characteristic length of the system and $\omega_\perp$ is transverse trapping frequency.

Introducing the scaled quantities $z\rightarrow z/\sqrt{{\hbar}/ {m\omega_\perp}}$, ~$t\rightarrow {t}{\omega_\perp}$, we can derive the GPE for a spin component of the system as

\begin{small}
\begin{align}
i\partial_{t}\Phi_{j}  = &\left[-\partial_{z}^{2}/2+
(-1)^j \left(i\kappa\partial_{z}-\delta/2\right) + V\left(z\right)\right]\Phi_{j}+\Omega\Phi_{3-j}\nonumber \\
& -\left[ \gamma \left| \Phi_{j}\right|^{2} + \beta \left|\Phi_{3-j}\right|^{2} + D_{j} + D_{3-j} \right] \Phi_{j},
\end{align}
\end{small}
with 
\begin{small}
\begin{subequations}
\begin{align}
&D_{j}=\mathcal F^{-1}\left\{ U_f\left(k_z,\theta\right) \cdot \mathcal F\left[\left|\Phi_{j}\left(z,t\right)\right|^{2}\right]\right\},\\
&U_{f}\left(k_{z},\theta\right)=\! D_{d}\!\left\{ \!1\!-3\cos^{2}\!\theta\!\left[1\!-\!G\!\left(k_{z}\right)\right]/2\!-\!3\!\sin^{2}\!\theta G\!\left(k_{z}\right)\!\right\}.
\end{align}
\end{subequations}
\end{small}

Here, $F(k_z)=\int \left| \Phi_{j} \left(z ,t\right) \right|^{2}e^{-ik_zz} dz$, $D_{d} = \mu_0 \mu^2/ 12 \pi L^2$ denotes the DDI strength.

\bibliography{apssamp}

\end{document}